\documentclass{an}
\usepackage{graphicx}
\usepackage{times}
\usepackage{fancyhdr}
\begin{document}

\title{Rotation and variability of young very low mass objects}

\author{Alexander Scholz$^1$, Ray Jayawardhana$^1$, Jochen Eisl{\"o}ffel$^2$, Dirk Froebrich$^3$}
\institute{$^1$ Department of Astronomy \& Astrophysics, University of Toronto,
60 St. George Street, Toronto, Ontario M5S3H8, Canada\\
$^2$ Th{\"u}ringer Landessternwarte Tautenburg, Sternwarte 5, D-07778 Tautenburg, Germany\\
$^3$ Dublin Institute for Advanced Studies, 5 Merrion Square, Dublin 2, 
Ireland}

\date{Received; accepted; published online}

\abstract{Variability studies are an important tool to investigate key properties of stars
and brown dwarfs. From photometric monitoring we are able to obtain information about rotation 
and magnetic activity, which are expected to change in the mass range below 0.3 solar masses, 
since these fully convective objects cannot host a solar-type dynamo. On the other hand, 
spectroscopic variability information can be used to obtain a detailed view on the accretion 
process in very young objects. In this paper, we report about our observational efforts to 
analyse the variability and rotational evolution of young brown dwarfs and very low mass stars.
}

\correspondence{aleks@astro.utoronto.ca}

\maketitle

\section{Introduction}

Variability is a key tool to study stellar properties. From simple photometric monitoring
alone it is possible to measure the rotation period, if the objects exhibit asymmetrically
distributed surface features, e.g. cool magnetic spots. The amplitude of the variability can be 
used to assess the properties of these spots, in particular if lightcurves in more than 
one filter are available (e.g. Bouvier \& Bertout 1989). Furthermore, accretion processes 
can be studied in detail based on spectroscopic time series, particularly by monitoring changes in
emission features like the H$\alpha$ line, which are produced in the hot accretion flow
(e.g., Johns \& Basri 1995). Until the late 1990s, most variability studies have focused on stars 
with masses $>0.4\,M_{\odot}$. They provided hundreds of rotation periods (see Stassun \& 
Terndrup 2003 for a recent review), detailed spot parameters for cool as well as for hot 
spots, and, for selected targets, a detailed view on the accretion process (e.g., Alencar \& 
Batalha 2002). For lower-mass objects, however, the observational database is still sparse. 
This motivated us to carry out a long-term project aimed to study the variability in the very 
low mass (VLM) regime, focusing on VLM stars and brown dwarfs (BDs), i.e. objects with masses 
below $0.4\,M_{\odot}$. Here we provide a summary of the outcomes of these studies. 

\section{Photometric rotation periods}

About 2000 photometric rotation periods have been measured for solar-mass stars over the 
past 25 years (Stassun \& Terndrup 2003). However, five years ago only a handful of 
periods were known for VLM objects (Mart\'{\i}n \& Zapatero Osorio 1997, Terndrup et al. 1999, 
Bailer-Jones \& Mundt 1999). Motivated by this lack of information, we and other groups have 
determined about 500 periods for VLM objects, where most of these periods have been measured 
in very young clusters (Herbst et al. 2001, Lamm et al. 2004, 2005). Although the VLM rotation
database is still sparsely populated in the BD regime with only about 30 periods known
for substellar objects (Scholz \& Eisl{\"o}ffel 2004a (SE04a), 2005 (SE05), Bailer-Jones 
\& Mundt 2001, Joergens et al. 2002, Zapatero Osorio et al. 2003), we are now able to carry 
out meaningful comparisons between solar-mass stars and VLM objects in terms of their 
rotational behaviour. Our own long-term project was dedicated to study the rotational 
evolution on timescales of a few 100 Myrs, and provided 80 periods for VLM objects with ages 
ranging from 3 to 700\,Myrs, increasing the number of periods in this age and mass range by 
a factor of 14. In the following, we will discuss mass-period relationship and rotational 
evolution mainly based on this sample.

\begin{figure}
\resizebox{\hsize}{!}
{\includegraphics[angle=-90]{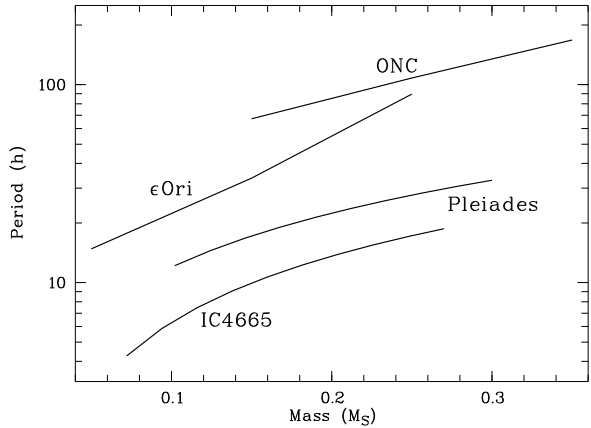}}
\caption{Average period vs. mass for objects in four open clusters. For IC4665 and the Pleiades, 
the lines show a linear fit to the available period data (Scholz 2004, SE04b). For $\epsilon$\,Ori 
(SE05) and the ONC (Herbst et al. 2001), the median period is plotted.}
\label{massper}
\end{figure}

In Fig. \ref{massper} we show average period vs. mass in four open clusters. The data for
$\epsilon$\,Ori, IC4665, and Pleiades are taken from our own project (see Scholz \& Eisl{\"o}ffel
2004 (SE04b), SE05, Scholz 2004), whereas the mass-period relationship for the ONC was derived 
by Herbst et al. (2001). In all clusters we see a similar trend: The average period clearly 
decreases with decreasing mass. This has two consequences: a) VLM objects rotate, on average, 
faster than solar-mass stars. b) In the VLM regime, the average period is correlated with the 
object mass. This trend appears to continue well-down into the substellar regime, with the 
result that the typical period of BDs is below one day. Since this period-mass relationship 
is already established in the youngest clusters (the ONC has an age of about 1\,Myr), it must 
have its origin in the earliest stages of the rotational evolution. Although an explanation 
for this effect is still missing, it might be an important signature of the VLM formation 
process.

\begin{figure}
\resizebox{\hsize}{!}
{\includegraphics[angle=-90]{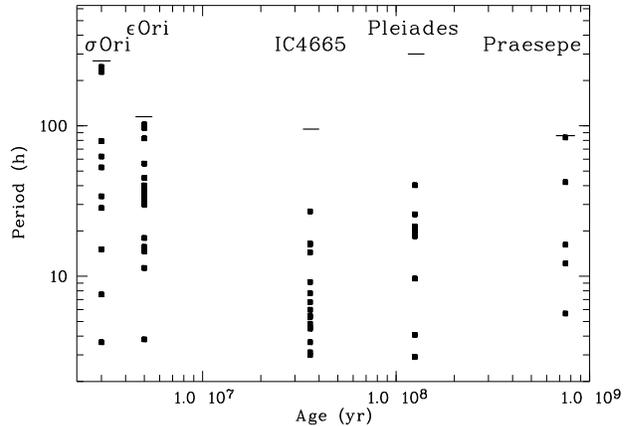}}
\caption{Rotation period of VLM stars in five open clusters as a function of age
(SE04a/b SE05, Scholz 2004). Horizontal lines show the detection limits.}
\label{rotevo}
\end{figure}

The rotational evolution can be analysed based on Fig. \ref{rotevo}, which shows the periods
for VLM stars from our monitoring campaigns in five open clusters as a function of age. The lower
period limit is, within the statistical uncertainties, more or less constant with age. Fast rotators
exist at all ages in the VLM regime, in clear contrast to solar-mass stars. On the other hand,
the upper period limit clearly evolves with time, it decreases from about 10 days at 3\,Myr
to 2 days at 125\,Myr, and increases again thereafter. We aimed for a deeper understanding
of this behaviour, and constructed a set of simple models, which take into account the basic
angular momentum regulation mechanisms (SE04b). The approach of the models was to use the
periods in our youngest cluster ($\sigma$\,Ori, 3\,Myr) as starting points and to evolve these
periods forward in time, to generate evolutionary tracks in the period-age diagram. The first
model, which includes only the contraction of the objects, can reproduce the rotational
acceleration between 3 and 100\,Myr, but fails to explain the subsequent rotational braking.
In particular, about half of the periods in the Pleiades and Praesepe are too long to
be explained by simple angular momentum conservation. Therefore, we included a wind braking
law in the model. It turned out that the classical Skumanich type braking ($P \propto \sqrt{t}$)
is too strong, and would produce periods longer than 100\,h at the age of the Pleiades, which,
according to our rotation periods and available $v\sin i$ data (Terndrup et al. 2000) do not
exist. Instead, we can reproduce most of our periods by using an exponential braking law
($P \propto \exp{(t)}$), which provides only weak angular momentum loss by stellar winds.
Please note that these conclusions are not affected by age uncertainties for the open clusters,
which may be substantial, particularly at very young ages.

The only objects which pose a problem for these models are the fastest rotating objects
with periods $<10$\,h at ages of 3-5\,Myr. In all models, these objects
would rotate with periods below the lower period limit in the Pleiades at the age of 125\,Myr.
For these ultra-fast VLM objects, we have to take into account that their breakup limit 
is between 3 and 5\,h. Therefore, we can expect that the fast rotation itself significantly
influences their evolution, e.g. by mass-loss due to strong centrifugal forces or strongly 
oblate shapes. It is therefore not surprising that our simple models fail to explain the
evolution of these objects. Clearly, more sophisticated modeling is required for objects
close to breakup. Recapitulating, we are able to explain the rotational evolution of most
VLM objects with ages between 3 and 700\,Myr by taking into account contraction and weak 
angular momentum loss by stellar winds.

\section{Spot properties}

The rotation period is not the only important property that can be obtained from photometric
monitoring. The amplitude of the variability can be used to assess the properties of the
surface features, which are responsible for the flux modulation. For young stars and BDs,
these surface features are believed to be magnetically induced cool spots. As we have shown
in SE04b, the amplitudes are reduced at least by a factor of 2.4 in the VLM regime (see also 
the discussion in Scholz, Eisl{\"o}ffel \& Froebrich 2005). In addition, we see evidence for 
a low rate of active, i.e. photometrically variable, objects in the VLM regime (SE04b). 
This is a clear indication for a change of the spot properties, roughly at a mass of 
$0.4\,M_{\odot}$. 

Small amplitudes can be caused by either a concentration of spots at polar latitudes,
small contrast between spots and photospheric environment, low spot coverage, or a more
symmetric spot distribution. To distinguish between these scenarios, Doppler imaging would
be desirable. Since this is still at the edge of the observational capabilities, we carried
out multi-filter monitoring to obtain first constraints on the VLM spot properties
(Scholz et al. 2005). A sample of VLM stars in the Pleiades was monitored simultaneously in the
I-, J-, and H-band, using two telescopes at the German Spanish Astronomical Centre on Calar
Alto/Spain.

\begin{figure}[t]
\resizebox{\hsize}{!}
{\includegraphics[angle=-90]{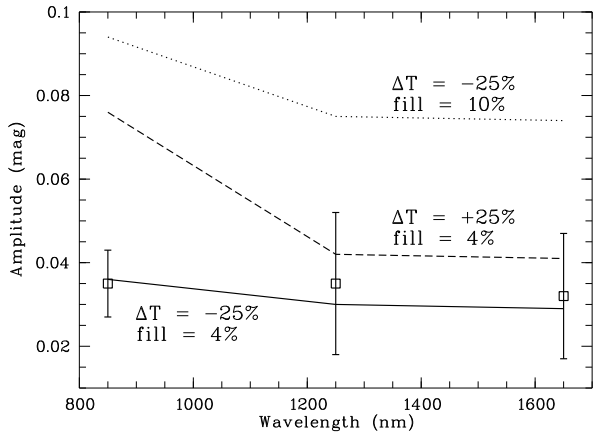}}
\caption{Photometric amplitude for BPL129, a VLM star in the Pleiades, as a function
of wavelength. Overplotted are model calculations for the given set of spot parameters.}
\label{spots}
\end{figure}

In Fig. \ref{spots} we show the results for a particular object, the star BPL129
(Pinfield et al. 2000) with an approximate mass of $0.15\,M_{\odot}$. Squares show the
observed lightcurve amplitudes, the lines are model calculations for given temperature difference
between spot and photosphere ($\Delta T$) and filling factor (fraction of the hemisphere
covered by spots). These models are based on theoretical spectra for VLM stars by
Allard et al. (2001), and assume an effective temperature of 3200\,K for BPL129.
The amplitudes in all three bands are comparable and best reproduced by models with 
$\Delta T = 20 \ldots 30$\% and filling factors $<5$\%. These are the first constraints 
of spot parameters for an object with such a low mass. The value for $\Delta T$ is 
comparable with similar measurements for solar-mass stars, but the filling factor is 
quite low when compared with values for more massive stars. The major downside of 
photometric monitoring is that we only observe asymmetrically distributed spots.
Thus, a low filling factor means that we see either low spot coverage or a symmetric
spot distribution. The low amplitudes in the VLM regime are thus probably a consequence
of either very few spots or a change in the spot distribution.

Is it possible to explain the change in the spot properties as well as the fast rotation
of VLM objects in the context of the fundamental properties of these objects? Maybe the most
significant difference between solar-mass stars and VLM objects is that the latter will never
develop a radiative core. Today it is widely believed that the large-scale magnetic field of 
the Sun is produced in the interface layer between convective and radiative zone by means of an 
$\alpha\omega$-type dynamo. The fully-convective VLM objects cannot have such a dynamo, because 
they have no radiative core. Since they are still magnetically active (see e.g. Mohanty \& Basri
2003), they must have an alternative type of dynamo. One suggestion is that VLM objects host
a turbulent dynamo, which works throughout the convection zone, and which should generate
only small-scale magnetic fields (Durney et al. 1993). If we assume that this model is adequate,
we expect fast rotating objects, because small-scale fields would cause inefficient angular
momentum loss by stellar winds. We also expect a change in the spot properties, probably to
a more symmetric spot distribution. Thus, our main results are consistent with a scenario
where VLM objects have only small-scale fields because they are fully-convective.

\begin{figure*}[t]
\includegraphics[width=5.5cm,height=6.7cm,angle=0]{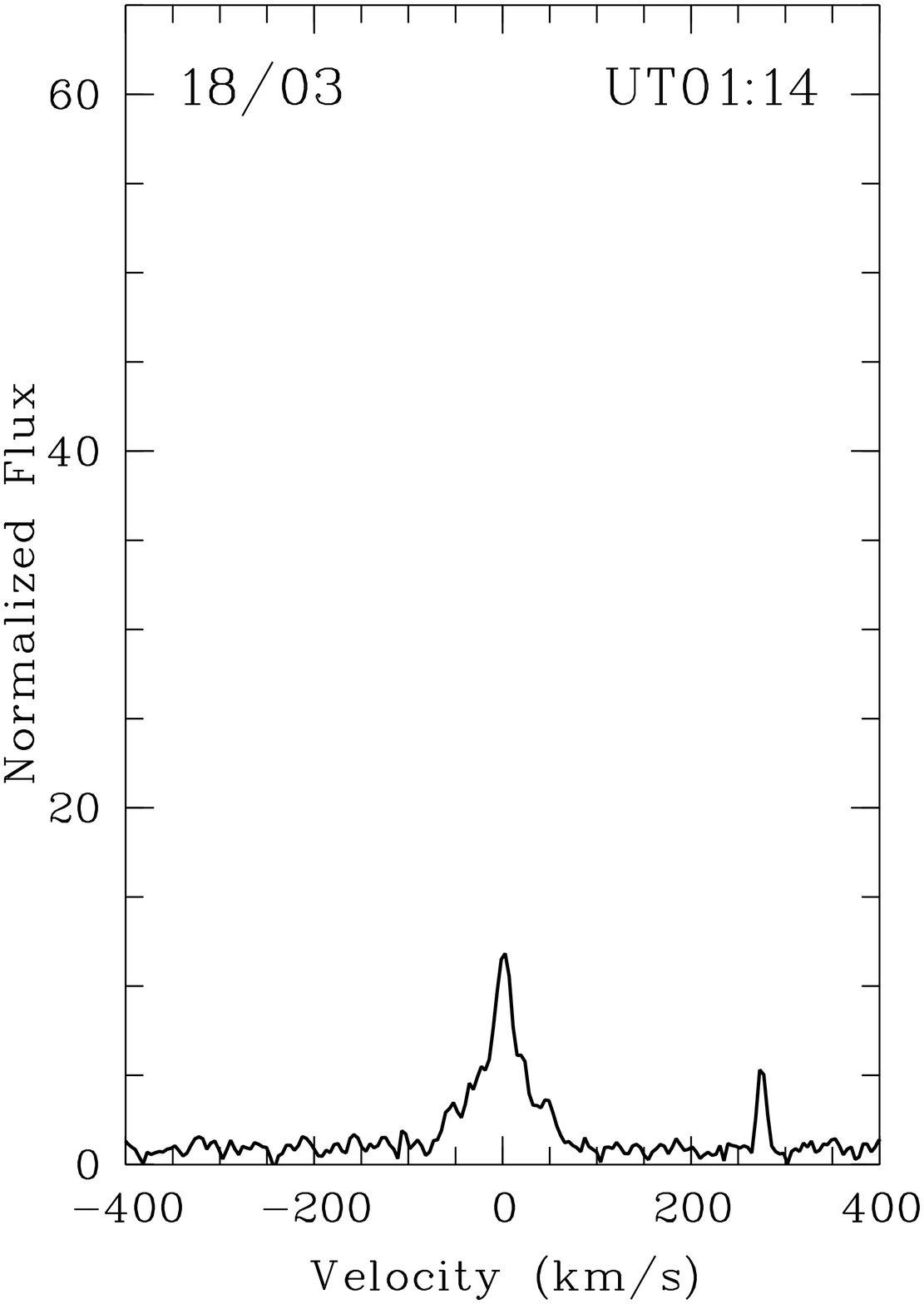} \hfill
\includegraphics[width=5.5cm,height=6.7cm,angle=0]{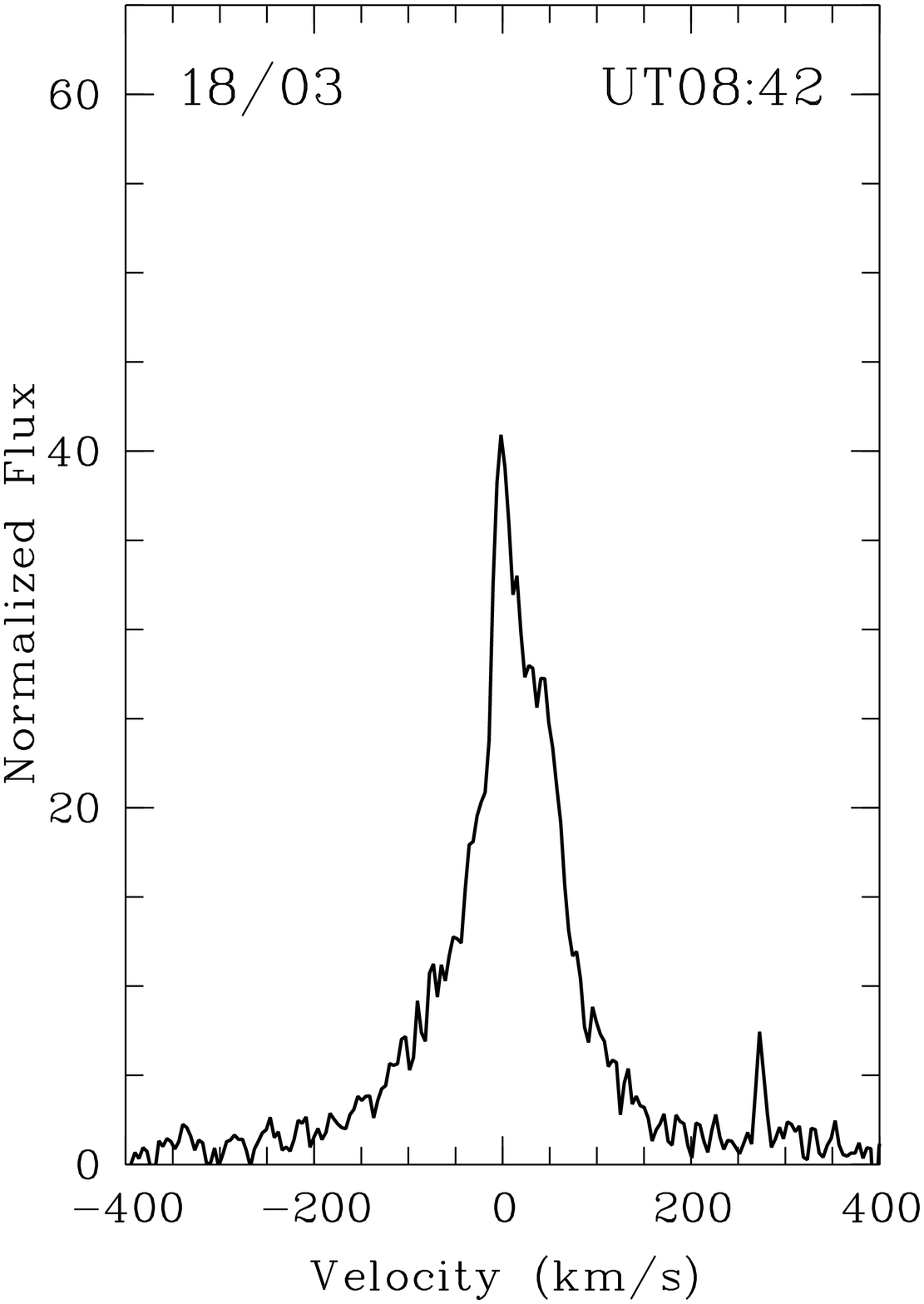} \hfill
\includegraphics[width=5.5cm,height=6.7cm,angle=0]{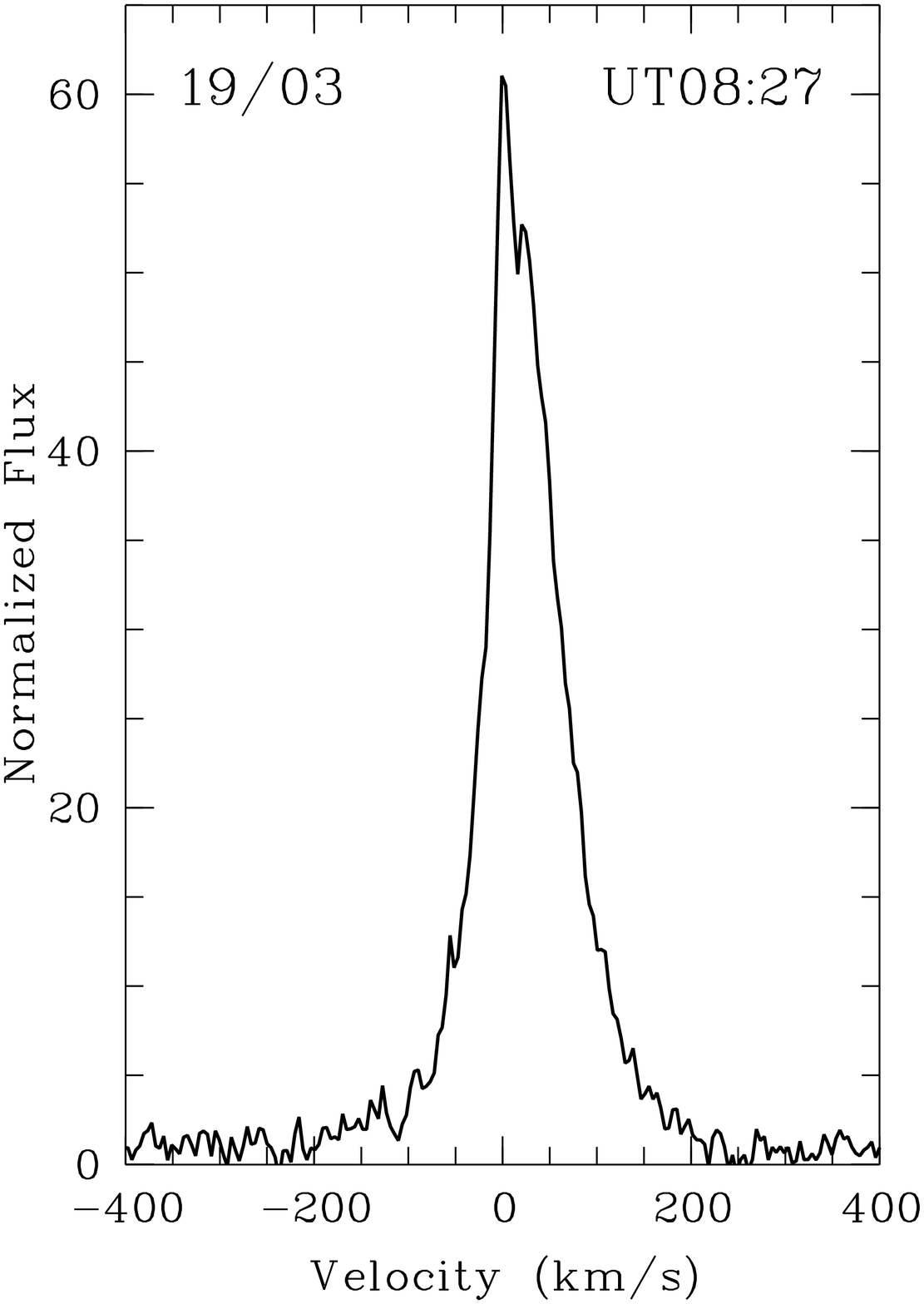}\\
\caption{Time series of the H$\alpha$ profile for the brown dwarf 2M1101. The 
sequence shows three spectra taken with separations of 8 and 24\,h hours. Emission 
features at $\sim 270\,\mathrm{km s^{-1}}$ are caused by non-perfect background 
subtraction.}
\label{2m1101}
\end{figure*}

\section{Monitoring accreting brown dwarfs}

In our photometric monitoring campaigns in very young clusters in Orion, we found not only
low-amplitude, strictly periodic lightcurves. In addition, we identified 11 objects showing 
high-amplitude variations with amplitudes up to 1\,mag and partly irregular variations
(SE04a, SE05). These highly variable VLM objects show lightcurves very similar to those of 
classical T Tauri stars, which are strongly variable because they accrete material from a 
circumstellar disk. The usual interpretation for their variability is the existence of
hot spots formed by the accretion flow, which co-rotate with the objects and therefore
cause high-amplitude variations. Instabilities in the accretion process can account for
irregular variability. Therefore, we explain the high-amplitude variations of VLM objects as 
a consequence of ongoing accretion. This interpretation is confirmed by near-infrared 
photometry and low-resolution spectroscopy: Highly variable objects tend to show infrared
colour excess and strong H$\alpha$ emission, indicative for the existence of an accretion
disk. Thus, we have identified a sample of accreting VLM objects by means of a variability
study.

Accreting objects show a very complex type of variability. To disentangle the contributing
effects, and to study hot spots, accretion rate variability, and disk geometry on VLM objects,
we carried out the first comprehensive spectroscopic monitoring campaigns for accreting young 
BDs. Here we show first results from high-resolution monitoring based on spectra taken with the
MIKE spectrograph at the Magellan/Clay 6.5-m telescope on Las Campanas in spring 2005. In
a first paper we report about dramatic changes in the H$\alpha$ profile and intensity for
the BD 2M1207 (Scholz, Jayawardhana, \& Brandeker 2005, Jayawardhana et al., this volume). 
To demonstrate that strong accretion rate variations are not unique in the substellar 
regime, we show in Fig. \ref{2m1101} a part of the H$\alpha$ time series for the BD 2MASS
J11013205-7718249, a likely member of the Cha I star forming region. On timescales 
of 1-2 days, the equivalent width in H$\alpha$ increases by a factor of six, and the 10\% 
width by a factor of two. This corresponds to a change in the accretion rate by about an 
order of magnitude (Natta et al. 2004). At the same time, alternative accretion indicators 
appear in the spectra, e.g. HeI and H$\beta$ emission, which are faint or not detected before 
this burst. In addition, the H$\alpha$ profile develops a significant red absorption 
feature, indicating infalling material. Thus, we probably witnessed a strong accretion 
burst on this object.

In total, we monitored six brown dwarfs with high-resolution spectroscopy, and all six 
targets are variable. Therefore, variability information is essential to assess the 
accretion properties of stars and BDs. As shown by Mohanty et al. (this volume), the
accretion rate is correlated with object mass, but the individual objects scatter by
one order of magnitude around this correlation. This accretion-mass relationship
might be explained by Bondi-Hoyle accretion (see Padoan, this volume), but in this
case we should expect much less scatter. Here we demonstrate that accretion rate 
variability might account for the largest part of the noise. By taking into account 
information about variability, we might be able to constrain correlations between
accretion rate and fundamental properties more precisely.

\section{Conclusions}

We report on a variety of variability studies of VLM stars and brown dwarfs, aimed 
at a deeper understanding of rotation, activity, and accretion for objects with masses 
$<0.4\,M_{\odot}$. In a long-term project, we measured 80 rotation periods 
for VLM objects with ages between 3 and 700\,Myr. At all evolutionary stages, we see a 
clear tendency of decreasing average periods with decreasing mass in the VLM regime. The 
rotational evolution can be understood by taking into account contraction and weak
angular momentum loss by stellar winds. From an analysis of the variability amplitudes
and multi-filter monitoring, we see evidence for a change in the properties of magnetic
spots at $\sim 0.4\,M_{\odot}$. Either VLM objects have a symmetric spot distribution or 
very low spot coverage in comparison with more massive stars. Accretion shows up in monitoring
studies as high-amplitude photometric variation and as strong variability in accretion-related
emission lines, in particular H$\alpha$. We demonstrate that VLM objects in some cases
show accretion rate variations up to one order of magnitude. Variability studies are
thus an important component in the investigation of substellar properties.

\acknowledgements
We would like to thank the organisers of the IAC/TNG workshop on ultralow-mass stars
for a lively and stimulating conference. Part of this work was supported by the German 
\emph{Deut\-sche For\-schungs\-ge\-mein\-schaft, DFG\/} project numbers 
Ei~409/11--1, Ei~409/11--2, as well as by an NSERC grant and University of 
Toronto startup funds to R. Jayawardhana.

\end{document}